\documentclass[conference]{IEEEtran}
\IEEEoverridecommandlockouts
\usepackage{cite}
\usepackage{amsmath,amssymb,amsfonts}
\usepackage{algorithmic}
\usepackage{graphicx}
\usepackage{caption}
\usepackage{subcaption}
\usepackage{textcomp}
\usepackage{xcolor}
\usepackage{amsmath}
\usepackage[utf8]{inputenc}
\usepackage{float}
\usepackage[utf8]{inputenc}
\usepackage[english]{babel}
\DeclareUnicodeCharacter{2212}{-}
\DeclareUnicodeCharacter{2191}{\uparrow}
\DeclareUnicodeCharacter{200E}{}
\usepackage{array, makecell}
\usepackage{boldline}
\usepackage{colortbl}
\usepackage{soul}
\usepackage{multirow}
\usepackage{multicol}
\def\BibTeX{{\rm B\kern-.05em{\sc i\kern-.025em b}\kern-.08em
    T\kern-.1667em\lower.7ex\hbox{E}\kern-.125emX}}

\begin{document}

\title{The Study of Optical Properties for Ordered and Disordered Silicon Nanowire Structures$^{\dagger}$\\
{\footnotesize \textsuperscript{$^{\dagger}$}All authors contributed equally in this work}

\thanks{D. Sarker acknowledges the financial support from the Bangladesh University of Engineering and Technology through the Postgraduate Fellowship program.}
}


\author{
\IEEEauthorblockN{Ohidul Islam}
\IEEEauthorblockA{\textit{Department of Electrical and Electronic Engineering} \\
\textit{Shahjalal University of Science and Technology}\\
Sylhet 3114, Bangladesh\\
ohidulislam4860@gmail.com}\\   

\IEEEauthorblockN{
K. B. M. Sharif Mahmood}
\IEEEauthorblockA{\textit{Department of Electrical and Electronic Engineering} \\
\textit{Shahjalal University of Science and Technology}\\
Sylhet 3114, Bangladesh\\
kbmsharifmahmood@gmail.com}
\and 

\IEEEauthorblockN{Dip Sarker*}
\IEEEauthorblockA{\textit{Department of Electrical and Electronic Engineering} \\
\textit{Bangladesh University of Engineering and Technology}\\
Dhaka 1205, Bangladesh\\
*0422062377@eee.buet.ac.bd}\\

\IEEEauthorblockN{
Joyprokash Debnath}
\IEEEauthorblockA{\textit{Department of Electrical and Electronic Engineering} \\
\textit{Shahjalal University of Science and Technology}\\
Sylhet 3114, Bangladesh\\
joyprokash025@gmail.com}
}


\maketitle
\thispagestyle{plain}
\pagestyle{plain}
\begin{abstract}
We designed ordered and disordered silicon (Si) nanowire structures and analyzed their optical performance using the finite-difference time-domain (FDTD) technique. We studied the orderness of nanowire structures by calculating scalar variance. This study reveals that utilizing disorder structures can increase the average absorbance of Si nanowire structures. Spatial electric field distributions provided insights into light-matter interaction, indicating that disorder structures had higher path lengths compared to the periodic structure. We achieved an average absorbance of 41.46\% for the hyperuniform Si nanowire structure with a maximum absorbance of 78.18\%. Intuitively, we obtained $\sim$70\% high absorbance compared to periodic Si nanowire structure. Our findings will be conducive to designing new efficient solar cells and photodetectors.
\end{abstract}
\section{Introduction}
Researchers have recently garnered interest in nanowire structures due to their unique optical properties that make them appealing for various applications, including photonics, sensing, imaging, and more~\cite{Raman_2023}. These properties arise from their size, shape, and material composition. Structures comprising nanowires have two ramifications depending on their arrangement: one is the periodic or ordered nanowire structure, and the other is the disordered nanowire structure. In optics, the arrangement of nanowires is precisely organized in periodic nanowire structures in order to obtain maximum efficiency. Therefore, this periodic arrangement is suitable for numerous optoelectronic devices, such as solar cells~\cite{Garnett}, LEDs~\cite{Lee}, photodetection~\cite{Zhu}, etc. Contrarily, the field of disordered nanowire optics has gained significant attention from scientists and academics due to its ability to offer equivalent or superior performance with flexible designs in some cases. Numerous investigations have been performed to examine such disorder structures experimentally and theoretically. These studies explored various applications and fundamental phenomena, including the behavior of light in disordered mediums~\cite{Wiersma1997}, the manipulation of whispering gallery resonators for directional control~\cite{Hassan}, the ability to adjust photonic bandgaps~\cite{Tahmid}, and efficient light management in solar cells~\cite{Pratesi}. However, is there any possibility of obtaining an ordered disorder structure so that new technology can emerge? 

Disordered hyperuniformity structures, nature's hidden order, are rare patterns in matter that suppress large-scale density fluctuations like an ordered structure in a small zone. Joe Corbo first observed this structure in Chicken's eye. Moreover, this ordered disorder structure behaves like a crystal without a Bragg peak in an intermediate to large length scale; however, it shows the properties of liquid in a small length scale. Salvatore Torquato worked for more than twenty years to intricate the mystery of this structure. They explored the photonic bandgap of hyperuniformity structures that helped to classify electronic and phononic bandgaps (PBG) in disordered materials~\cite{Torquato_2009}. Harnessing this PBG, waveguides~\cite{MM}, waveguide polarizers~\cite{Wen}, graded effective index materials~\cite{Yi}, and topological insulators~\cite{NP} have been reported utilizing this unique structure. However, to the best of our knowledge, a comprehensive comparison of optical properties among these structures has yet to be explored. 

\begin{figure*}[htbp] 
\centering\includegraphics[width=18 cm]{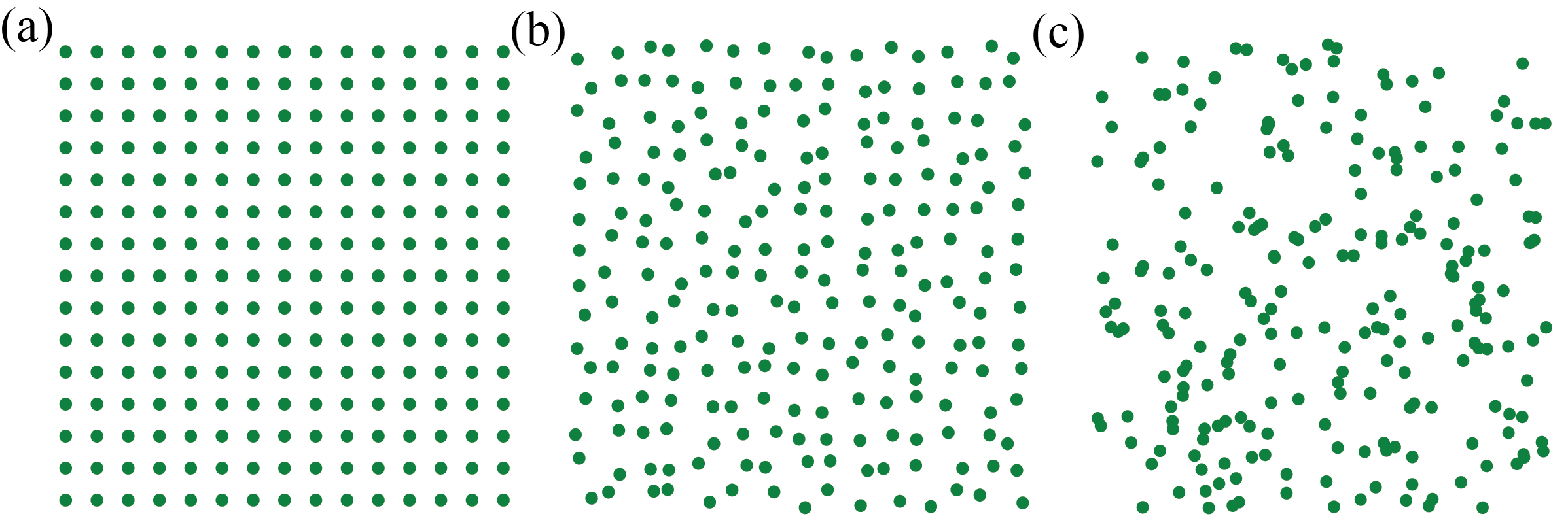}
\caption{Schematic illustration of (a) periodic, (b) hyperuniform, and (c) random silicon nanowire structure. Here, the xy-cross-sectional views are depicted.}
\label{fig:design}
\end{figure*}
In this work, we designed three Si nanowire structures and analyzed their optical properties employing the FDTD technique. Initially, we optimized the structural parameters of structures with scalar variance and nanowire's counts with respect to distance. Afterward, we adopted power monitors to calculate reflectance, transmittance, and absorbance spectra. The electric field reveals the light-matter interaction in the designed nanowire structures. Moreover, we provided a comparative table of the optical performances of these structures. Our designed structures not only pave the way for designing ingenious photonic structures but also provide insight into how disorder structures can mimic periodic structures.    
\section{Design and Methodology}
\begin{figure}[ht!] 
\centering\includegraphics[width= 8 cm]{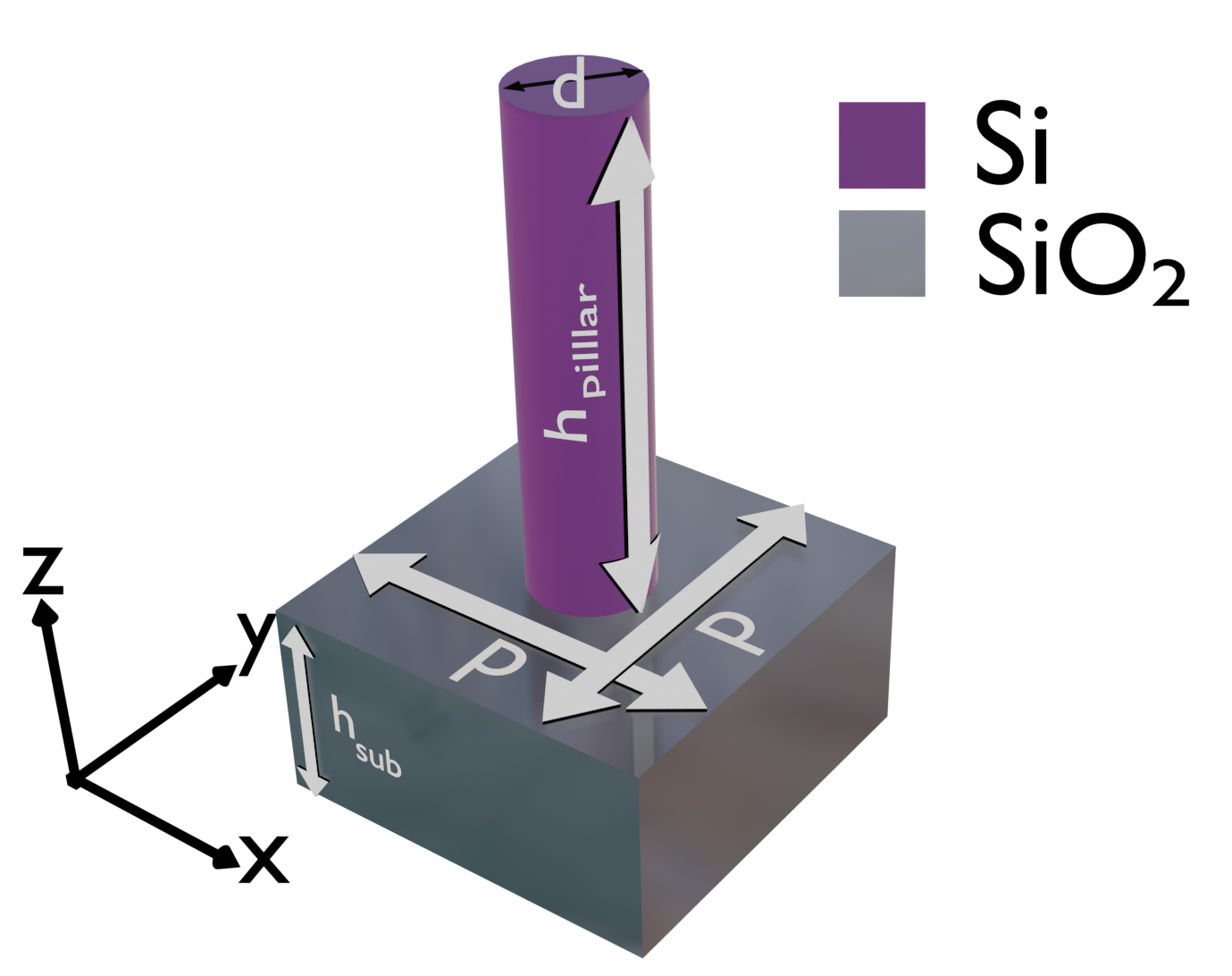}
\caption{Schematic illustration of a unit cell Si nanowire from the periodic structure where we utilized SiO$_2$ as a substrate layer. Here, d, h$_{\mbox{pillar}}$, h$_{\mbox{sub}}$, P$_x$, and P$_y$ are nanowire diameter, pillar height, substrate thickness, period of x- and y-directions, respectively.}
\label{fig:unit}
\end{figure}

Si in solar cells is a widely used material because of its advanced fabrication techniques and facilities; however, the absorbance of Si is poor. Hence, Si solar cells require a few microns of thickness to trap light efficiently~\cite{DS, Nelson}. We aimed to obtain absorbance as much as possible by employing light-matter interactions so that light experienced more path length and was absorbed in a small-thickness material. Therefore, we designed three Si nanowire periodic, hyperuniform, and random structures and performed their optical performances employing the three-dimensional (3D) FDTD technique. Figure~\ref{fig:design} delineates the two-dimensional (2D) representation of Si nanowire periodic, hyperuniform, and random structures from the xy-plane. We used SiO$_2$ as the substrate layer due to its adhesiveness to Si so that low lattice mismatch and high efficiency can be obtained. The 3D FDTD technique is a time-consuming method and requires high computational facilities. Therefore, we employed periodic boundary conditional in the x- and y-directions for a unit cell of periodic nanowire structure. The schematic of a unit cell is depicted in Fig.~\ref{fig:unit} where d, h$_{\mbox{pillar}}$, h$_{\mbox{sub}}$, P$_x$, and P$_y$ are nanowire diameter, pillar height, substrate thickness, period of x- and y-directions, respectively. We considered d of 160 nm, h$_{\mbox{pillar}}$ of 800 nm, h$_{\mbox{sub}}$ of 160 nm, P$_x$ of 250 nm, and P$_y$ of 250 nm for our unit cell simulation.  
As hyperuniform and random structures are aperiodic in the x- and y-directions, the dimensions of the structures were set to be 6 $\mu$m in the x- and y-directions, with a lattice constant, a, of 400 nm. We adopted a perfectly matched layer (PML) of these two structures' x- and y-directions. In order to generate the hyperuniform structure, a small degree of randomness was introduced to the positional arrangement of the periodic structure. The spatial coordinates (x + $\delta$x, y + $\delta$y) of nanowires inside the hyperuniform structure may be determined by using the following equation: 
\begin{equation}
\begin{aligned}
    \delta x &= a \times R_i + r_x \times P_x, \\
    \delta y &= a \times C_i + r_y \times P_y.
\end{aligned}
\label{eqn:hyper-eqn}
\end{equation}

Here, $R_i$, and $C_i$ represent the $i$-th row and column numbers, respectively. $r_x$ and $r_y$ denote the uniformly produced random values for the x- and y-axis, respectively, and the interval of [0, a] was maintained. P\textsubscript{x} and P\textsubscript{y} are the percentages of randomness parameters in both the x and y-axis. During the computation, the value of P\textsubscript{x} and P\textsubscript{y} was set to 60\%. The hyperuniform structure's randomness can be regulated according to the expectations by varying P\textsubscript{x} = P\textsubscript{y}. Fig.~\ref{fig:design}(b) illustrates the hyperuniform structure. Contrarily,  the controlled kind of randomness disappeared as the random structure was created. Fig. \ref{fig:design}(c) shows the distribution of Si nanowires in the random structure. This range of x = $\mathrm{\left[0, width \right]}$ and y = $\mathrm{\left[0, height \right]}$ was maintained while generating the desired number of nanowires. We used 225 Si nanowires for our hyperuniform and random structures where the d, h$_{\mbox{pillar}}$, and h$_{\mbox{sub}}$ remained the same. We employed 12 PML layers in the z-direction for all structures. A reflectance and a transmission monitor were utilized to analyze the optical properties of the designed structures. Absorbance was calculated by,
\begin{equation}
\begin{aligned}
    A(\lambda) = 1-R(\lambda)-T(\lambda).
\end{aligned}
\label{eqn:A}
\end{equation}
Here, $A$, $R$, and $T$ denote the absorbance, reflectance, and transmittance, respectively. The temperature was set to be 300 K for our study. 
\section{Result and Discussions}
When the scalar variance of a many-particle system within an observation window increases more slowly than the volume of the window, the many-particle system is said to be hyperuniform. Moreover, this property can be observed for a many-particle periodic structure. Hence, we calculated the scalar variance to analyze periodic, random, and hyperuniformity structures. For this calculation, we varied the observation window radius and enumerated the scalar variance, $\sigma^2(R)$, by,
\begin{equation}
\begin{aligned}
    \sigma^2(R) = \langle N^2(R)\rangle - \langle N(R) \rangle ^2.
\end{aligned}
\label{eqn:SV}
\end{equation}
Here, $N(R)$ is the number of nanowires in the observation window radius, $R$.  
\begin{figure}[ht!] 
\centering\includegraphics[width=8 cm]{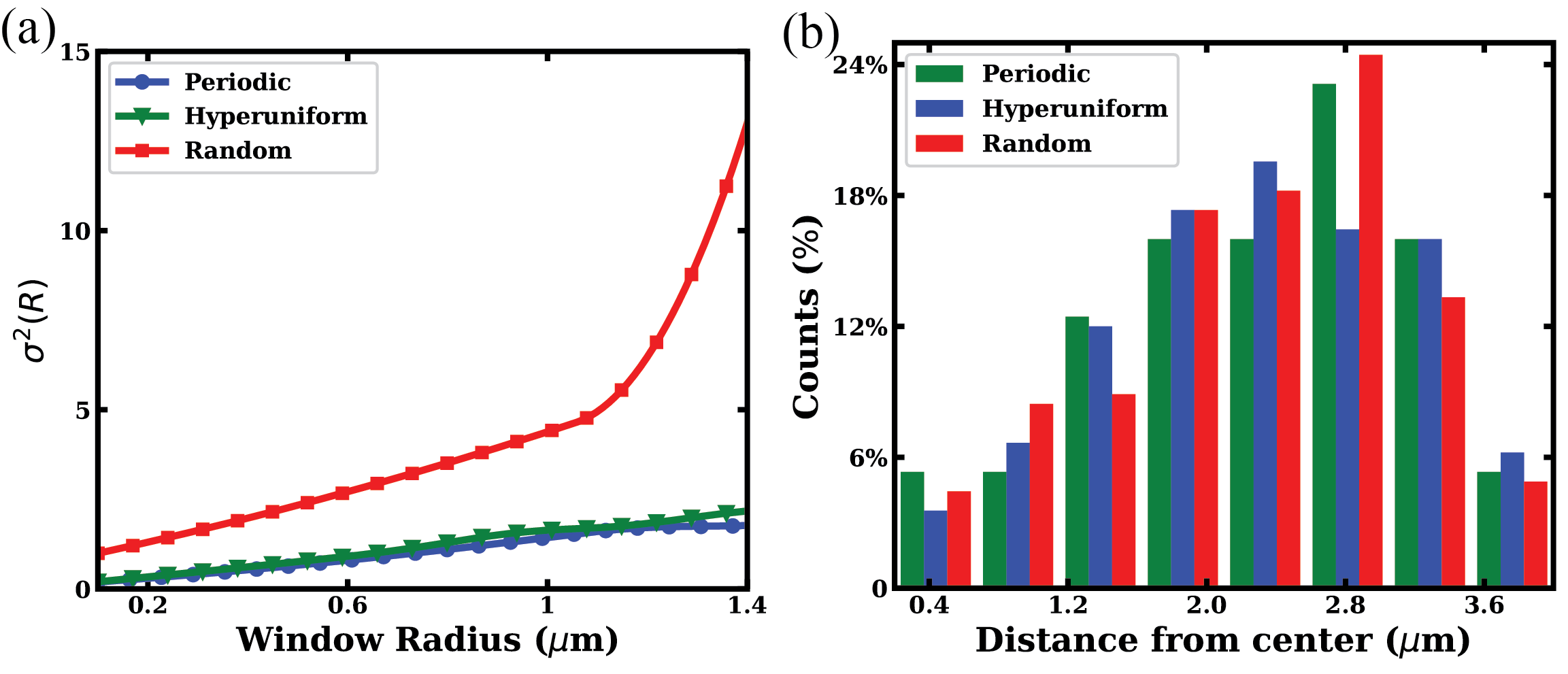}
\caption{(a) Scalar variance of three structures for a circular window located inside the structures. (b) Distribution of Si nanowire's count by changing distance from the center for all structures.}
\label{fig:variance-distribution}
\end{figure}
Though hyperuniform structure distributed Si nanowires arbitrarily, the $\sigma^2(R)$ of hyperuniform structure changed slowly compared to the window radius and followed the $\sigma^2(R)$ of periodic structure due to its unique nanowires distribution, as shown in Fig.~\ref{fig:variance-distribution}(a). Hence, hyperuniform structure is an ordered disorder many-particle system. However, random structure exhibited no such consequences. It can be inferred from Fig.~\ref{fig:variance-distribution}(a) that the $\sigma^2(R)$ of random structure became steeper with increasing window radius. Moreover, we calculated the percentages of Si nanowires by varying distances from the center of the designed structures. The calculated Si nanowire counts reveal why the $\sigma^2(R)$ of hyperuniform structure follows the $\sigma^2(R)$ of periodic structure. Figure \ref{fig:variance-distribution}(b) provides the Si nanowire counts of designed structures with varying distances. Our designed hyperuniform structure abode the Si nanowire distribution of periodic structure.   
\begin{figure*}[ht!] 
\centering\includegraphics[width=18 cm]{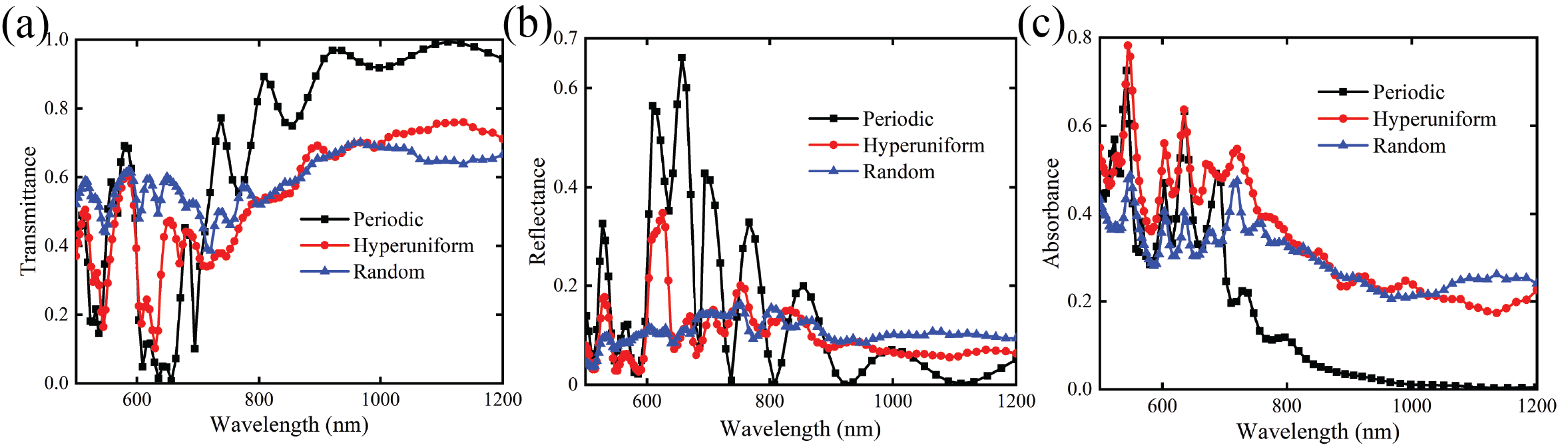}
\caption{(a) Transmittance, (b) Reflectance, and (c) Absorbance spectra of periodic (black line), hyperuniform (red line), and random (blue line) structures. We obtained more absorption in hyperuniform structure compared to other structures.}
\label{fig:ART}
\end{figure*}

Fig.~\ref{fig:ART} shows the transmittance, reflectance, and absorbance spectra for periodic, hyperuniform, and random Si nanowire structures. Periodic structure exhibited high transmittance of around 60\% at 720 nm. Moreover, the transmittance increased with increasing wavelength, as can be seen in Fig.~\ref{fig:ART}(a). A maximum transmittance of 0.9931 was obtained at the wavelength of 1110 nm. In the shorter wavelength region, the periodic Si nanowire structure experienced the highest reflectance compared to other structures, as depicted in Fig.~\ref{fig:ART}(b). Thus, we obtained an average absorbance of 0.2501 for the periodic structure, which was substantially poor due to increasing reflectance and transmittance at shorter and higher wavelength regions, respectively. Interestingly, the random Si nanowire structure experienced a constant average transmittance, reflectance, and absorbance in the entire wavelength region. 

\begin{figure*}[ht!] 
\centering\includegraphics[width=18 cm]{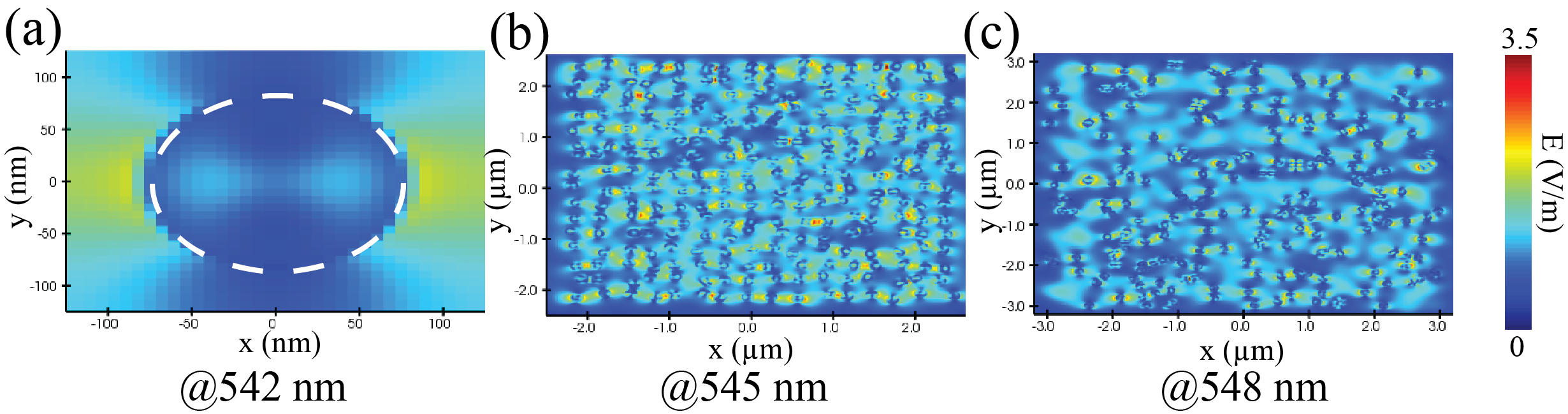}
\caption{Spatial xy-electric field distribution for (a) a unit cell of periodic, (b) hyperuniform, and (c) random structures at the absorbance peak of 542, 545, and 548 nm, respectively. The white dot circle in Fig.~\ref{fig:field}(a) depicts the nanowire of Si, and a color bar is provided in the inset.}
\label{fig:field}
\end{figure*}

Hyperuniform and random Si nanowire structures increased the path length of the light so that light experienced enough time to trap it in the Si nanowire structures~\cite{Nelson}. Hence, we achieved high average absorbances of 0.4146 and 0.4027 for hyperuniform and random Si nanowire structures, respectively, which was $\sim$70\% high absorbance compared to periodic Si nanowire structure. For better comprehension, the spatial electric field distributions are delineated for the periodic, hyperuniform, and random Si nanowire structures in Fig.~\ref{fig:field}. We observed high light-matter interaction in hyperuniform and random Si nanowire structures compared to the periodic Si nanowire structure where the hyperuniform structure experienced the most and exhibited the highest average absorbance. An extensive comparison of optical performance parameters for periodic, hyperuniform, and random Si nanowire structures is enlisted in Table~\ref{Tab:compare}.   
\renewcommand{\arraystretch}{1.5}
\begin{table}[htbp]
\caption{Comparative optical performance analysis of periodic, hyperuniform, and random structures}
\begin{center}

\begin{tabular}{c c c c c c c}
\hline
Structure & T$_{\mbox{max}}$ & R$_{\mbox{max}}$ &  A$_{\mbox{max}}$ & T$_{\mbox{avg}}$ & R$_{\mbox{avg}}$ & A$_{\mbox{avg}}$\\
\hline
Periodic & 0.9931 & 0.6612 & 0.7255 & 0.5747 & 0.1752 & 0.2501 \\

Hyperuniform & 0.7601 & 0.3475 & 0.7818 & 0.4903 & 0.1070 & 0.4146 \\

Random & 0.7014 & 0.1620 & 0.4875 & 0.5726 & 0.1031 & 0.4027 \\ 
\hline
\end{tabular}
\label{Tab:compare}
\end{center}
\end{table}
\section{Conclusion}
In this work, we designed three Si nanowire structures and extensively analyzed their optical properties using the FDTD technique. We calculated the scalar variance, which revealed the hyperuniformity structure imitated the nanowire distribution of the periodic structure. Moreover, we studied the absorbance, reflectance, and transmittance spectra for periodic, hyperuniform, and random Si nanowire structures. We obtained 70\% more absorbance for disorder structures compared to the periodic structure. Disorder structures exhibited strong light-matter interaction compared to the periodic structure, resulting in longer path lengths and high absorbance. Our study paves the way for designing novel photonics and optoelectronics devices, including solar cells, waveguides, topological insulators, etc.

\vspace{12pt}

\end{document}